\documentstyle[12pt]{article}

\newcommand{\text}{\rm}

\begin{document}

\author{{\bf Carla Goldman \footnotetext{{\bf e-mail:} carla@gibbs.if.usp.br}
and Robson Francisco de Souza} \\
Departamento de F\'\i sica Geral, \\
Instituto de F\'\i sica\\
Universidade de S\~ao Paulo, 
Caixa Postal 66318\\
05389-970 S\~ao Paulo, SP, Brazil\\
{\bf e-mail:} carla@gibbs.if.usp.br
}
\title{A Suggested Model for the Origin\\
of Mosaic Structure of DNA}
\date{September, 1996}
\maketitle

\begin{abstract}
In view of the critical behavior exhibited by a statistical model describing
a one dimensional crystal containing isotopic impurities and the fact that
for impurity mass parameter 3 times the host mass the critical temperature
can attain very high values, we suggest that the mosaic structure displayed
by certain nucleic acids (DNA) of eukaryotic organisms might have had its
origins in a phenomenon of condensation of {\it codons} ocurred at prebiotic
conditions. An appropriated map of that model onto some of the DNA features
allows one to predict power law behavior for both correlation functions and
exons size distribution in binary sequences of nucleotides which are
distinguished by their protein coding or noncoding functions. Preliminary
studies of these quantities performed on intron-containing sequences from 
{\it GenBank} are presented here.
\end{abstract}

\newpage\ 

\section{Introduction}

Nucleotide sequences on DNA of high organisms display a mosaic structure in
such a way that expressed sequences ({\it exons}) are interrupted by
intervening sequences ({\it introns}) which do not code for aminoacids \cite
{exin}. Up to the present, the role of introns are not unraveled although
there are beliefs that their presence on the eukaryotic genome promote
biological stability since mutations on noncoding regions are not expected
to affect information contents regarding protein synthesis \cite{livrao}.

Uncertainties on whether introns have been either introduced into or
withdrawn from embryonic sequences of nucleotides have raised doubts on the
common assumption that procaryotic genome is primitive relatively to the
eukaryotic genome. Up till now, the dispute between introns-early \cite
{early} and introns-late \cite{late} hypotheses appears to \thinspace be
unresolved.

It is worth to notice in this respect that there are evidences favoring the
presence of exons of all sizes on eukaryotic sequences and the existence of
a close relationship between exons distribution along such sequences and
protein structure \cite{proteinre}. The question of whether such a
distribution has a prebiotic origin is then closely related to the
introns-early/introns-late dilemma \cite{origin}.

In the last few years there have been numerous efforts to rationalize the
genetic information stored in DNA nucleotide sequences from a statistical
viewpoint \cite{stvosli}. Apart from differences of methodology in using 
{\it GenBank} data base, these works present numerical information on the
two-point autocorrelation function $C(r)$ in one-dimensional space (or its
Fourier transform power spectrum), for nucleotide separation distance $r$
(in appropriate units) on a same DNA strand. The essential question
addressed in these works relates to disclose long-range correlations (or,
equivalently, power-law behavior) in these sequences by distinguishing
nucleotides according to their pyrimidine (Cytosine, Thymine) or purine
(Adenine , Guanine) base content . Criteria based on the number of
hydrogen-bonds linking complementary nucleotides in pairs have also been
used.

In this context, the relevance of the coding and noncoding parts of
eukaryotic sequences for interpreting the genetic information of DNA has
been upraised since S. Nee's' suggestion that the ultimate source of
long-range correlations reported in early studies \cite{stvosli}, could be
the underlined mosaic structure of genome \cite{nee}. This observation has
led to further studies on purine and pyrimidine organization along separate
coding and noncoding sequences. Nevertheless, the question on whether the
mosaic structure of the genome is responsible for signaling long range order
in purine-pyrimidine distribution, remains under debate \cite{stanley93} .

From a physical viewpoint one of the interests of studying statistical
properties of the genome resides on the fact that, by finding long- range
behavior for $C(r)$, one might address the origin of nucleotide base
sequences organization to a critical phenomenon that eventually occurred
during biological evolution. In this respect, one can find in the literature
nonequilibrium statistical mechanics models proposed to describe the
dynamics of nucleotide sequences under the action of some natural
evolutionary driving processes \cite{othermodel}.

We choose to address questions related to the genome organization by
distinguishing nucleotides not between purines and pyrimidines but instead,
regardless their base content, by distinguishing nucleotides according their
loci on an exon or intron region of a DNA sequence . We believe that direct
information on statistical properties of mosaic structure shall be helpful
for tracing the origins and functions of introns, which is our main concern
here.

The focus of our studies on nucleotide sequences was entirely suggested by
the results obtained for equilibrium properties of a statistical mechanics
model proposed recently to investigate conditions for isotopic order (or
``isotopic fractionation'' ) on a harmonic crystal chain containing isotopic
impurities \cite{eu}. On the basis of an appropriated map (see below), we
argue that this model is also suitable for describing some features of DNA
concerning, in particular, its mosaic structure.

The relevant aspect of the original model equilibrium thermodynamics we want
to emphasize here is that at temperatures below a certain critical
temperature $T_C$, it displays a condensed phase in which impurities
aggregate into clusters of all sizes. Within this phase, the cluster
distribution function and two-point auto-correlation functions are expected
to exhibit temperature dependent power-law decay \cite{fifel}. Moreover, it
was also found in Ref. \cite{eu} that for masses $m_a$ of host particles and 
$m_b$ of impurities satisfying the particular relationship 
\begin{equation}
\label{cod}m_b\sim 3m_a 
\end{equation}
the critical temperature for condensation can attain very high values \cite
{nota}.

Making a parallel of this model with a similar description for DNA molecule,
allows us to conjecture that the origin of DNA mosaic structure might have
been due to a {\it codon }\cite{codon} condensation into clusters (or {\it %
exons}) that eventually occurred during prebiotic synthesis of some nucleic
acids. Within this picture, we suggest that thermodynamical stability of DNA
with respect to small spatial fluctuations of its molecular components had a
crucial role for establishment of an ordered state (mosaic structure) in
which codons appear segregated from protein noncoding regions. Important to
stress is that the present proposal do not relate to the emergence of
particular sequences of purine and pyrimidine bases; it is restricted to the
mosaic features, regardless base content.

In Section 2 we review the model of Ref. \cite{eu} and extend the results in
order to estimate the asymptotic behavior of related pair correlations
functions and clusters size distribution. On the basis of a map of this
model onto DNA nucleotide sequences suggested in Section 3, we show some
numerical results for these quantities obtained in preliminary studies using
data from {\it GenBank}. A discussion is in Section 4.


\section{A model Hamiltonian}

Here we review the original model of Ref. \cite{eu} and explore some of its
statistical properties to describe aspects of systems of interest.

\subsection{Isotopic order : phonon induced interactions}

As mentioned, the original lattice model Hamiltonian describes a one
dimensional crystal containing isotopic impurities interacting by harmonic
potential 
\begin{equation}
\label{hamilt}H=\sum_{i=1}^N\{\frac{m_i}2\ \stackrel{\cdot }{u}_i^2+\frac
K2u_i(u_i-u_{i-1})\} 
\end{equation}
where $u_i$ and $m_i$ are respectively, the displacement from equilibrium
position and mass of the particle at lattice site $i$. $K$ is a common force
constant. There are present in this system two isotopic species, A (host)
and B (impurity), with masses $m_a$ and $m_b$, respectively. By introducing
site-dependent spinlike variables $\sigma _i$ which assume the value 0 (for
host species) and 1 (for impurity), it is possible to write $m_i$ as: 
\begin{equation}
\label{mi}m_i=m_a+(m_b-m_a)\sigma _i 
\end{equation}

Now, it is assumed that this system can be driven into a region of
sufficiently high temperatures where the particles would freely interchange
positions with each other. The nature of such processes is immaterial here,
conditioned that their characteristic times are very small if compared to
typical lattice relaxation times. The question posed in Ref. \cite{eu} is
whether in this situation the phonons of the lattice could induce positional
correlations among impurities leading to some sort of isotopic order at
thermodynamical equilibrium. For that, we introduce phonon creation $\left(
b_q^{\dagger }\right) $and phonon annihilation $\left( b_q\right) $operators
as follows:

\begin{equation}
\label{cre}\stackrel{\cdot }{u}_j=\sum \left[ \frac 1{2Nm_a\omega _q}\right]
^{\frac 12}\left( b_q+b_{-q}^{\dagger }\right) e^{iqj} 
\end{equation}
and 
\begin{equation}
\label{anih}u_j=\sum \left[ \frac 1{2Nm_a\omega _q}\right] ^{\frac 12}\left(
b_q-b_{-q}^{\dagger }\right) e^{iqj} 
\end{equation}
rewrite $H$ in (\ref{hamilt}) as

\begin{equation}
\label{hamilt2}H=\sum_q\omega _qb_q^{\dagger }b_q+\frac \alpha
N\sum_{qq^{\prime }}\widetilde{\sigma }_{q-q^{\prime }}(\omega _q\omega
_{q^{\prime }})^{\frac 12}B_qB_{q^{\prime }} 
\end{equation}
where $B_q=b_q-b_{-q}^{\dagger },$%
\begin{equation}
\label{alpha}\alpha =\frac 14\left( m_b/m_a-1\right) , 
\end{equation}
\begin{equation}
\label{sigma}\widetilde{\sigma }_k\equiv \sum_j\sigma _je^{ikj} 
\end{equation}
and $\omega _{q\ }=\left[ 2K(1-\cos q)/m_a\right] ^{1/2}$ is the dispersion
relation for free phonons for%
$$
q\in \{-\pi ,(\frac{-N-2}N)\pi ,...,\frac{-2\pi }N,0,\frac{2\pi }N,...\pi
\}. 
$$

To approach equilibrium properties of this system, we consider its grand
partition function at fixed temperature $\beta ^{-1}:$

\begin{equation}
\label{particao1}\Xi (\beta ,\mu )=\sum_{\{\sigma _j\}}\left(
\sum_{\{n_q\}}e^{-\beta H}\right) \,e^{\beta \mu \sum_j\sigma _j} 
\end{equation}
where the sums extend over all spins configurations $\{\sigma _j\}$ and over
phonon occupation number in the $q$ mode $\{n_q\}.$ $\mu $ is a chemical
potential that controls the density of impurities and $H$ is as in (\ref
{hamilt2}). Integration over phonon variables allows one to write $\Xi $ as 
\begin{equation}
\label{particao2}\Xi (\beta ,\mu )=Z_0\sum_{\{\sigma _j\}}e^{-\beta \Delta
F(\{\sigma _j\})}\,e^{\beta \mu \sum_j\sigma _j} 
\end{equation}
where 
\begin{equation}
\label{en livre}\Delta F(\{\sigma _j\})\equiv -\beta ^{-1}\ln \frac{%
\sum_{\{n_q\}}\left\langle \{n_q\}\left| e^{-\beta H}\right|
\{n_q\}\right\rangle }{\sum_{\{n_q\}}\left\langle \{n_q\}\left| e^{-\beta
H_0}\right| \{n_q\}\right\rangle }\equiv 
\end{equation}
$$
\equiv -\beta ^{-1}\ln Z/Z_0 
$$
is an effective interaction among impurities induced then by the lattice
phonon contents. $Z_0$ is a normalization constant defined to set $\Delta
F(0)=0.$

We've found that, for a given spin configuration and low impurity density, $%
\Delta F$ is separated into a sum 
\begin{equation}
\label{fator}\Delta F=\sum_l\Delta F_{l^{(r)}} 
\end{equation}
such that $\Delta F_l^{(r)}$ is the effective energy of an isolated cluster
of impurities (indexed by $r$) containing $l$ components; namely, a $l$%
-cluster \cite{isolated}.

This enables us to focus on an isolated $l$-cluster and extract from $\Delta
F_l$ a cluster ''surface energy'' term $E_l$ whose asymptotic behavior is
given by 
\begin{equation}
\label{el}E_l\sim \frac{(K/m_a)^{1/2}}{\pi (2-\Lambda )^2}\ln l 
\end{equation}
as {\it l} goes to infinity, with 
\begin{equation}
\label{lambda}\Lambda =\frac{m_b+m_a}{m_b-m_a} 
\end{equation}
and a cluster bulk energy $l\Phi $, in agreement to the model studied in
Ref. \cite{fifel}. Accordingly, 
\begin{equation}
\label{deltaf}\Delta F_l\sim l\Phi +E_l 
\end{equation}

In the following we make further considerations to discuss the nature of the
phase transition exhibited by this model on the basis of equations (\ref
{fator}) and (\ref{deltaf}).


\subsection{Clusters Size Distribution}

It is our purpose here to characterize the phase transition displayed by the
model above in the limit of very low impurity density. For this, we replace
the sum over spin variables in (\ref{particao2}) by a sum over all
configurations of isolated impurity clusters of all sizes and positions.
Within this approximation we treat the system as an ideal lattice gas
mixture of different molecular species.

Let $a_l$ be the number of {\it l}-clusters in a given configuration of the
system. Then (\ref{fator}) becomes 
\begin{equation}
\label{deltaf2}\Delta F\sim \sum_{l=1}^\infty a_l(l\Phi +E_l) 
\end{equation}
and (\ref{particao2}) can be approximated by 
\begin{equation}
\label{particao3}\Xi (\beta ,\mu )\simeq Z_0\prod_l\sum_{a_l=0}^\infty \exp
[-\beta a_l(l\Phi +E_l)].\frac{(Lz_l)^{a_l}}{a_l!} 
\end{equation}
where $L$ is the chain length and $z_l=e^{\beta \mu _l}$ is the {\it l}%
-cluster activity with $\mu _l$ being the corresponding chemical potential.
At thermodynamical equilibrium, 
\begin{equation}
\label{mul}\mu _l=l\mu 
\end{equation}

Defining a {\it l}-cluster {\em internal partition function }$q_l$ as 
\begin{equation}
\label{interpar}q_l=\exp [-\beta (l\Phi +E_l)] 
\end{equation}
expression (\ref{particao3}) reads

\begin{equation}
\label{particao4}\Xi (\beta ,\mu )\simeq Z_0\prod_{l=1}^\infty e^{Lq_lz_l} 
\end{equation}
from which the average number of l-clusters per unit length $\rho _l(\beta
,\mu )$ can readily be obtained: 
\begin{equation}
\label{rol}\rho _l(\beta ,\mu )=\lim _{L\rightarrow \infty }\frac
1Lz_l\left( \frac{\partial \ln \Xi }{\partial z_l}\right) _{L,T}=l^{-\frac
\beta {\beta _c}}.\exp [-\beta l(\Phi -\mu )] 
\end{equation}

In expression (\ref{rol}), we have used relation (\ref{mul}) and defined a
critical inverse temperature $\beta _c$ by

\begin{equation}
\label{tc}\beta _c=\left[ \frac{(K/m_a)^{\frac 12}}{\pi (2-\Lambda )^2}%
\right] ^{-1} 
\end{equation}

On summing (\ref{rol}) over all values of {\it l}, we obtain the average
number of clusters of any kind per unit length; namely, 
\begin{equation}
\label{nu}\nu (\beta ,\mu )=\sum_{l=1}^\infty \rho _l(\beta ,\mu
)=\sum_{l=1}^\infty l^{-\frac \beta {\beta _c}}.[e^{-\beta (\Phi -\mu )}]^l 
\end{equation}
so that the fraction of {\it l}-clusters at thermodynamical equilibrium, or
equivalently, the probability of finding a {\it l}-cluster in the system, is
given by 
\begin{equation}
\label{pl}P_l(\beta ,\mu )=\frac{\rho _l}\nu 
\end{equation}

Let us now review the properties of the sum in (\ref{nu}) \cite{grad}:

(a) If $e^{-\beta (\Phi -\mu )}<1,$ then $\nu $ converges for $\beta >0;$

(b) If $e^{-\beta (\Phi -\mu )}=1,$ then $\nu $ converges for $\beta >\beta
_c$ or $T<T_c=1/k_B\beta _c.$

The last condition expresses the critical behavior of the model which
becomes apparent by studying the limit 
\begin{equation}
\label{limpl}\lim _{\mu \uparrow \Phi }P_l(\beta ,\mu )=\lim _{\mu \uparrow
\Phi }\frac{l^{-\frac \beta {\beta _c}}}{\nu (\beta ,\mu )}. 
\end{equation}
If $T>T_c$ then $\lim _{\mu \uparrow \Phi }\nu (\beta ,\mu )$ $\rightarrow
\infty $ so that for temperatures higher than $T_c$ the probability of
finding stable clusters of relative large sizes is negligible. However, this
limit attains finite values in regions where $T<T_c.$ Consequently, at
sufficiently low temperatures the statistical properties of the system are
governed by the L\'evy distribution 
\begin{equation}
\label{pllim}P_l(\beta >\beta _c,\Phi )\sim l^{-\frac \beta {_{\beta
_c}}}\equiv l^\eta 
\end{equation}
with 
\begin{equation}
\label{eta}\eta =\eta (\beta )=-\beta /\beta _c 
\end{equation}
which characterizes its condensed phase. Within this phase macroscopic
regions of impurities can be found at all scales.

We should notice that result (\ref{pllim}) which was obtained here for
noniteracting clusters, is expected to hold for this class of models even in
more general cases as those studied in Ref. \cite{fifel}.


\subsection{Pair Correlation Functions}

In addition to the size distribution we can also make quantitative
predictions on the behavior of impurity pair correlation functions which in
the present context are defined by 
\begin{equation}
\label{corr}C(\left| r\right| )=\left\langle \sigma _i\sigma
_{i+r}\right\rangle -\left\langle \sigma _i\right\rangle \left\langle \sigma
_{i+r}\right\rangle 
\end{equation}
for $\left| r\right| $ measuring distances between two impurity particles
along the considered chain. The set of variables $\{\sigma _i\}$ are
assigned according to (\ref{mi}). $\left\langle ...\right\rangle $ indicates
averages over statistical ensemble of the system.

Since the only contribution to the truncated function (\ref{corr}) comes
from configurations in which the considered two points are covered by a
single cluster we can use the approach introduced in the last Section and
estimate the behavior of $C(\left| r\right| )$ at the condensed phase by the
probability of finding any cluster of size greater or equal than $\left|
r\right| .$ Using result (\ref{pllim}), we then disclose asymptotic
power-law behavior 
\begin{equation}
\label{limcorr}C(\left| r\right| )\sim \sum_{l=\left| r\right| }^\infty
P_l(\beta >\beta _c,\Phi )\sim \left| r\right| ^\varepsilon 
\end{equation}
with 
\begin{equation}
\label{epslon}\varepsilon =\varepsilon (\beta )=1-\beta /\beta _c 
\end{equation}
as expected \cite{fifel}, \cite{fisher}. We remark that $1-\beta /\beta _c<0$
within this phase.


\section{Modeling Eukaryotic Sequences}

\subsection{Map onto DNA}

It is our intention in this Section to suggest a map of the model reviewed
above in order to describe aspects of DNA eukaryotic sequences. For this
purpose, it is relevant to notice in Eq. (\ref{lambda}) that for $\Lambda
\sim 2,$ i.e., for 
\begin{equation}
\label{codon}m_b\sim 3m_a 
\end{equation}
the critical temperature of condensation $T_c$ can attain arbitrarily large
values [see Eq. (\ref{tc})]. This result leads us to make some conjectures
concerning coding and noncoding organization along some nucleotide sequences:

(i) Let us assume that small positional displacements of nucleotides with
respect to their equilibrium positions along DNA $\alpha $-helix backbone,
which in turn generate phonons, are relevant degrees of freedom of the
macromolecule and can be accounted by the lattice model Hamiltonian in Eq.(%
\ref{hamilt}). For this, we assume also that the two DNA complementary
strands can be represented by a single chain of harmonic oscillators placed
along that direction. Accordingly, we shall consider that any two
nucleotides linked by H-bonds in complementary pairs are, in certain
conditions, constrained to move along chain direction as a single particle
of mass (Fig 1a):

\begin{equation}
\label{mat}m+m_A+m_T=m_{AT} 
\end{equation}
or 
\begin{equation}
\label{mcg}m+m_C+m_G=m_{CG} 
\end{equation}
where $m_{A(T,C,G)}$ is the mass of an Adenine (Thymine, Cytosine or
Guanine) base and $m$ is the mass of remaining sugar and phosphate
components of nucleotides.

$$
\text{FIGURES }\; \;\text{1(a) } \; \;,\text{1(b) } 
$$

Due to the fact that either complementary pair AT or CG comprise a
two-carbon ring purine base and a one-carbon pyrimidine joint to attain a
three-ring base pair (see Fig.1b), we can ascribe within reasonable
approximation the mass of a host particle $m_a$ defined in the model for
isotopes, to the mass of any combination A-T or C-G, i.e. 
\begin{equation}
\label{host single}m_a\sim m_{AT}\sim m_{CG} 
\end{equation}

(ii) In addition, we consider the possibility that eventually in our DNA
lattice model there are present some triplets of consecutive complementary
pairs of nucleotides which are also constrained (by action of mechanisms
external to the system) to move as single particles; each of these comprises
then six nucleotides. The number of such assembled triplets is controlled by
a chemical potential $\mu .$

In view of (\ref{host single}) we can then ascribe a mass $m_b$ of
``impurities'' particles to any of these triplets in such a way that
relation (\ref{codon}) holds, i.e. 
\begin{equation}
\label{mb}m_b\sim c_1m_{AT}+c_2m_{CG} 
\end{equation}
for $c_1,c_2$ integers assuming any value of the interval $[0,3]$ but
subjected to the condition $c_1+c_2=3.$

(iii) Finally, the assembled triplets are to be identified here with {\it %
codons} (accompanied by their complementary nucleotides), which are the
units of genome contents information.

In analogy to the model of last Section, we assign site dependent spinlike
variables$\ \left\{ \sigma \right\} $ to the lattice ordered points such
that 
\begin{equation}
\label{sigma1}
\begin{array}{lll}
\sigma _i=1, & \text{if} & i \;
\text{ is \; a \; codon } \\  & \text{or} &  \\ 
\sigma _i=0, & \text{if} & i \;\text{ is \; not \; a \; codon} 
\end{array}
\end{equation}
Notice that by this procedure, each codon occupies a single lattice site.
Notice also that, according to our purposes, the defined set $\{\sigma \}$
for chosen nucleotide sequences disregard differences concerning the nature
of purine-pyrimidine base pairs.

Steps (i), (ii) and (iii) accomplish the map. Let us examine some of its
consequences.


\subsection{Distribution of Coding Sequences: Mosaic Structure}

We are now able to make some quantitative predictions on the structure of
eukaryotic DNA sequences with respect to the distribution of coding regions.

According to the results of Sec. 2.2, and the above map, we expect that the
size distribution of coding regions along eukaryotic gene sequences, which
we refer here as DNA mosaic structure \cite{mosaic}, follows the power-law (%
\ref{pllim}) with $\beta /\beta _c>1$, in analogy to the distribution of
clusters of isotopic impurities of the physical model at condensed phase.

In order to get support for this conjecture we select sequences from {\it %
GenBank} and check for this property. We focus on Saccharomyces chromosomes
(SCCHRIII, SCCHRIX, YSCCHRVIN) which are among the longest intron-containing
sequences available at that data bank. We notice that such sequences were
selected from available{\it \ }data upon requirement of presenting complete
coding as well as noncoding regions (CDS sequences).

Figure 2 shows histograms where there are depicted the diverse sizes of the
respective coding regions. To extract representative functions for these
diagrams in each case we test both polynomial and exponential fittings which
are also shown.

$$
\text{FIGURE} \;\;\text{ 2} 
$$


\subsection{Codon Pair Correlation Functions}

In the same context we examine the behavior of codon pair correlation
functions. Figure 3 shows the variation of $C(r)$ with codon pair distance $%
r $ for various intron-containing nucleotide sequences taken from {\it %
GenBank}. These curves have been obtained upon assignments (\ref{sigma1}),
and assuming ergodicity of the systems of interest so that for each
sequence, we replace the ensemble averages indicated in (\ref{corr}) by
``temporal'' averages 
\begin{equation}
\label{corr5}C(\left| r\right| )=\frac 1L\sum_{i=1}^L\sigma _{i+r}\sigma
_i-\left( \frac 1L\sum_{i=1}^L\sigma _{i+r}\right) \left( \frac
1L\sum_{i=1}^L\sigma _i\right) 
\end{equation}
where the sequence length $L$ is measured in nucleotide units.

Fig. 3 shows our results for $C(r)$ obtained by evaluation of expression (%
\ref{corr5}) on the same DNA sequences studied last Section and considering
periodic boundary conditions. 
$$
\text{FIGURE} \;\; 	\text{ 3} 
$$
Comments on these results are in Sec.4.


\section{Discussion}

The above ideas relate to the origin and function of genome mosaic
organization. They are based on the results obtained previously for
equilibrium properties of a one-dimensional harmonic lattice chain model of
two isotopic components (masses $m_a$ and $m_b$) which was proposed recently
for studying isotopic fractionation. In favorable thermodynamic conditions,
this model exhibits a condensed phase in which \thinspace segregation takes
place, giving rise to macroscopic regions of just one component \cite{eu}.
In order to use these results to pursue a description of DNA mosaic
structure, we focus on the multiscale characteristic predicted for the
distribution of such aggregates and on the fact that the critical
temperature for phase transition can attain very high values for a
particular set of values for model parameters (i.e. for $m_b\sim 3m_a$)$.$
In this case, the condensed phase should be stable in a broad range of
temperatures including, in particular, our present room temperatures.

On the basis of the map of Section 3, we are then lead to suggest: (i) that
the mosaic structure might have had its origins in a condensation phenomena
of {\it codons} that took place at prebiotic conditions, and (ii) this
mosaic structure, being {\em thermodynamically }${\em stable}$ with respect
to the distribution of coding and noncoding sequences, might in fact be
related to the{\em \ biological} {\em stability} of the genome of high
organisms since it is believed that the presence of noncoding sequences
ensures relative low probabilities of harmful mutations on the genome \cite
{livrao}.

In order to find support to the model and to check for the plausibility of
related assumptions, we explore the fact that both exon size distribution
functions and codon correlation functions are predicted to display power-law
decay according to expressions (\ref{pllim}) and (\ref{limcorr})
respectively. It is remarkable in this respect that the decay power in both
cases are predicted to be temperature dependent. This fact shall be of some
relevance from a biological point of view, for it provides information on
the thermodynamics of gene formation.

We select sequences from {\it GenBank} for which calculation of these
quantities were performed. The results obtained are shown in Figures 2 and
3. In these calculations we distinguish nucleotides according to their loci
on a coding or noncoding region of each sequence, in agreement to our
assumptions.

Let us focus on the results obtained for pair correlations of selected
chromosome sequences, and restrict the analysis to the range of two-point
distances depicted in Figure 3. We notice that correlation curves for all
chosen sequences are smooth only in a relative small range, exhibiting in
the remaining an oscillatory behavior due, probably, to the absence of
experimental information on coding and non-coding regions in all
representative scales \cite{scale}. Even so, we observe that the decay
profile of the curves in Figure 3 can in principle be fitted by either
exponential or polynomial functions. Although the characteristic decay
parameter in each of the exponentials shown is very small (i.e., not greater
than $O$ $(10^{-3}),$ see Table 1) we should not, on the basis of these
results alone, discard the possibility of exponential in favor of polynomial
fittings. In face of this, the situation then appears to be inconclusive
concerning long-range order of coding regions in eukaryotic sequences of
nucleotides.%

\newpage

$$
\begin{tabular}{|c|c|c|c|c|c|}
\hline 
Sequence & $ m^{\exp }$ & $n^{\exp }$ & $\varepsilon ^{\exp }$ & $\eta ^{\exp
}$ & 
 
$\varepsilon ^{\exp }-\eta ^{\exp }$ \\  
\hline \hline
\text{SCCHRIII} & $-1.4\times 10^{-3}$ &$ -2.1\times 10^{-3}$ & $-.79 $&
$-1.83$  & 
$1.04$ \\  
\hline
\text{SCCHRIX} & $-7.7\times 10^{-4}$ & $-6.2\times 10^{-4}$ & $-.99 $& $-2.15
$ & 
$1.16 $\\  
\hline
\text{YSCCHRVIN} & $-8.3\times 10^{-4}$ & $-2.6\times 10^{-3}$ & $-.91$ & $-1.68$ & 
$.77  $\\
\hline
\end{tabular}
$$

{\it Table 1 }- Results of polynomial fitting exponents for exons size
distribution, $\eta ^{\exp },$ and codon correlation functions, $\varepsilon
^{\exp }$ (Fig.2 and 3 respectively), along the saccharomyces chromosome
nucleotide sequences indicated. Experimental information on coding and
non-coding regions along each of these sequences were taken from {\it GenBank%
} data bank. $m^{\exp }$ and $n^{\exp }$ are the corresponding results for
exponential fitting coefficients. Last column shows the difference $\eta
^{\exp }-\varepsilon ^{\exp }$ obtained in each case. According to our
theoretical predictions, this difference is expected to be of order one. 

\bigskip
			\bigskip

Let us now focus on the numerical results obtained for exons size
distribution. The fact that there are present in the histograms many gaps in
different positions corroborates our conjectures above on the oscillatory
behavior of correlation functions. Figure 2 presents both exponential and
polynomial fittings for the referred histograms. We observe that also in
this case either fittings seem to be adequate although, as for the case of
correlations, the decay parameter of each of the fitted exponentials is very
small.

Additional evidences in favor of polynomial fittings can nevertheless be
obtained in case of distribution functions. In fact, as can be observed in
Figure 2, rare events, namely, relative large coding regions present in each
sequence can only be accounted by polynomial fitting curves with $\beta
/\beta _c>1$, in agreement to our predictions.

Moreover, if we compare the decay profiles of related correlation and
distribution polynomial fittings we find that the differences between the
corresponding decay coefficients for all of the sequences studied is of
order one, as predicted by our model [compare expressions (\ref{pllim}) and (%
\ref{limcorr}) with results in the last column of Table 1).

In conclusion, although our numerical studies lack more information on other
scales of coding and non-coding regions of genome, it is still possible to
obtain strong support for the use of the model of Section 2 to describe
aspects of DNA mosaic structure, on the basis of a joint analysis of
available experimental data on both exons size distribution and codon
correlation functions.

\smallskip\ 

Finally, we comment on the hypothesis of Section 3. In our opinion, it is
important for the model credibility to find reasons at least for some of the
stated assumptions, which we review here: a) in our DNA lattice model, some
triplets of nucleotides complementary pairs are constrained to behave as
single particles concerning their interactions with the remaining system, in
contrast to other pairs which are not constrained in this way; b) in special
conditions (prebiotic), all particles can interchange positions with each
other.

One possibility to think on these is to imagine triplets being assembled at
the required prebiotic conditions, by the action of external agents that are
able to promote aggregation of any three consecutive nucleotide
complementary pairs at random positions on the lattice. Details of physical
mechanisms involved in this kind of process are immaterial here, since it is
also assumed that it happens in a in a very short time scale if compared to
the lattice relaxation time. Within this picture, the coding regions of DNA
would emerge due to a possible sequential action of these external agents.
According to the results presented here, stabilization of macroscopic
sequences of such triplets though, would be favorable only at regions of
temperatures below $T_C.$


\section{Acknowledgments}

This work was supported by Conselho Nacional de Desenvolvimento Cient\'\i
fico e Tecnol\'ogico (CNPq), Brazil.

C.G. acknowledges D.H.U Marchetti for illuminating discussions and for
reading the manuscript.



\section{Figure Caption}

\begin{description}
\item  {\bf Figure 1a: }Cytosine-Guanine (CG) and Adenine-Thymine (AT)
nucleotide complementary pairs. Notice that either CG or AT comprises three
aromatic rings.

\item  {\bf Figure 1b: }Single chain mass-string representation of a DNA
nucleotide sequence.

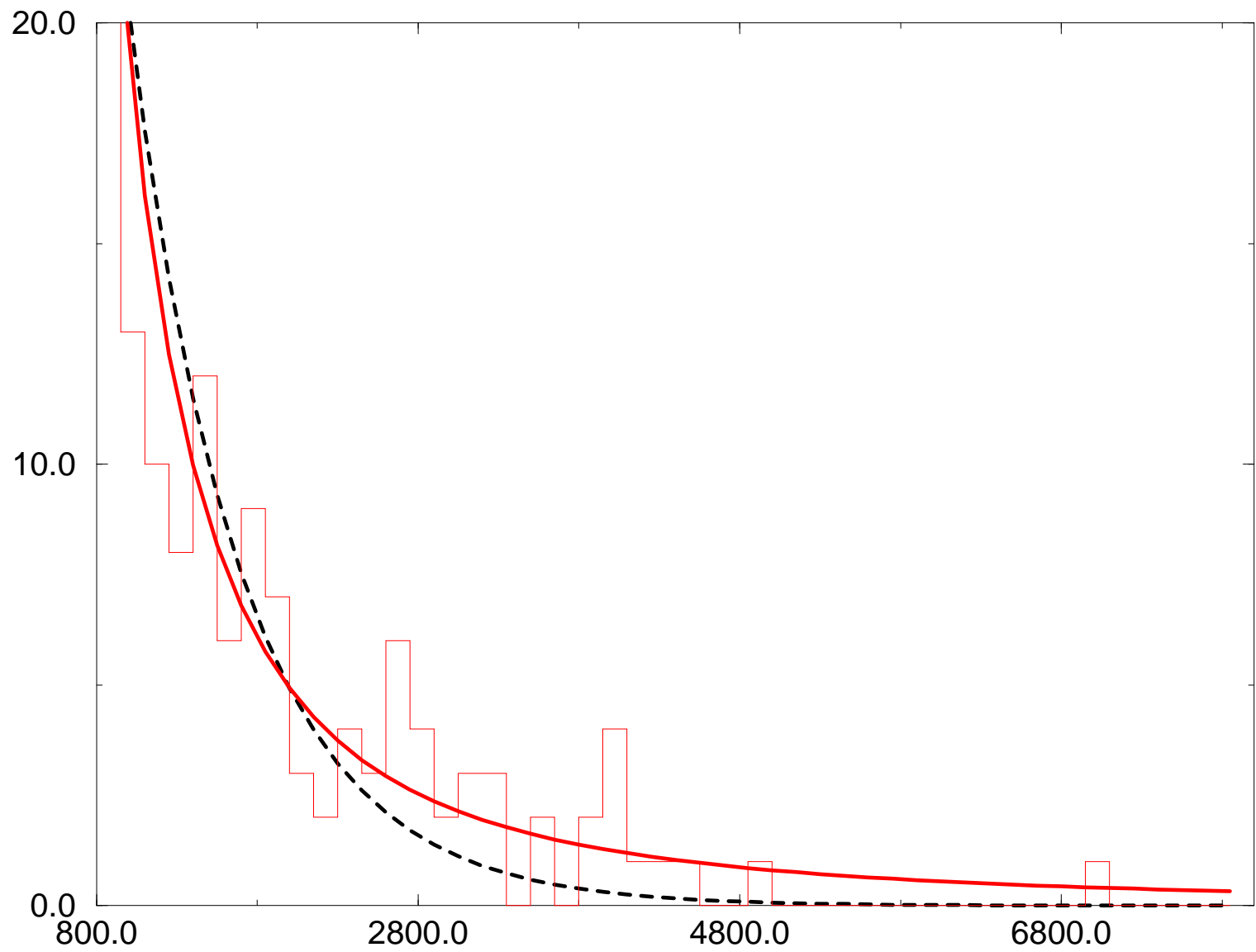
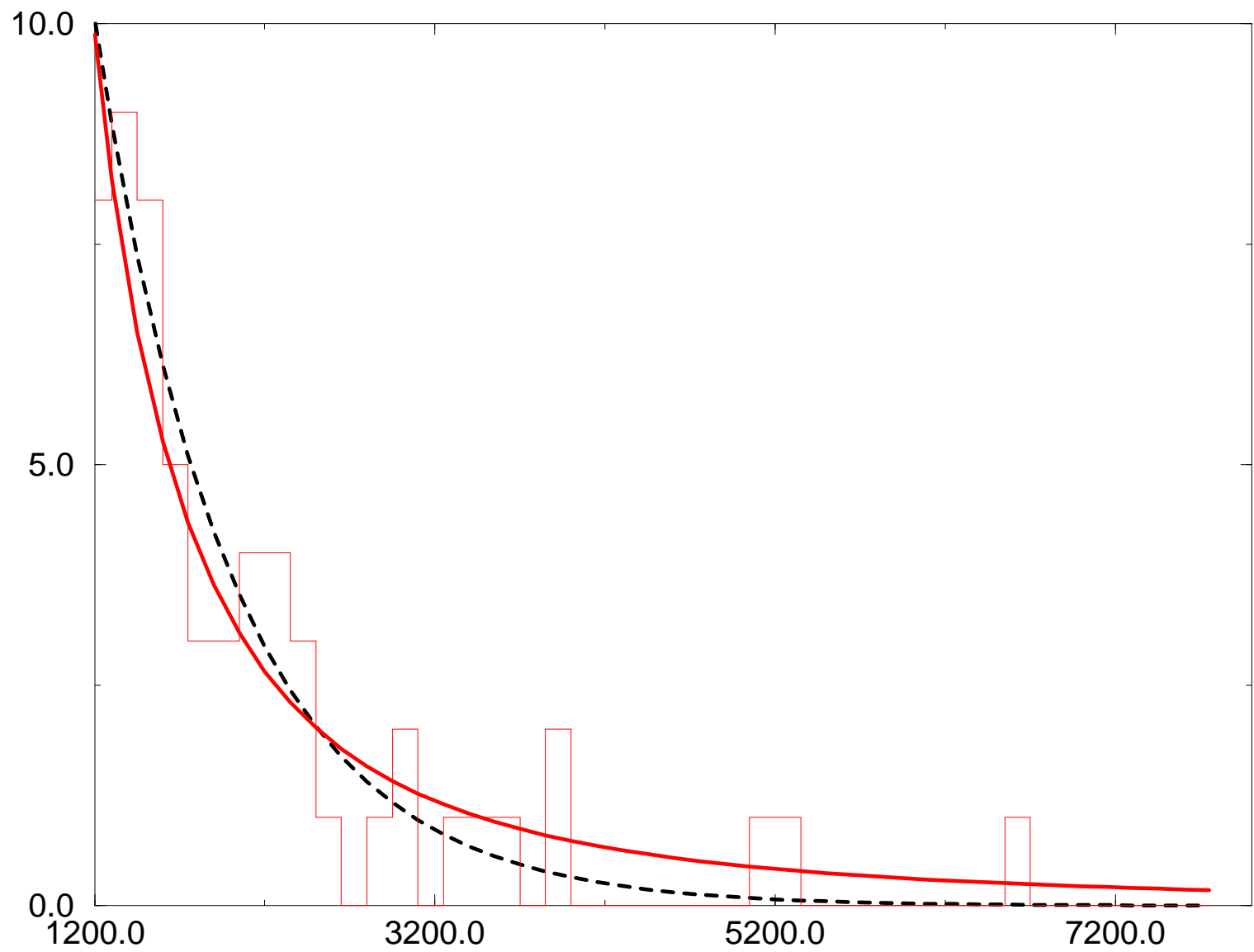
\item  {\bf Figure 2: }Exon size histograms for saccharomyces chromosome
nucleotide sequences taken from {\it GenBank} : (a) SCCHRIII [315,341 base
pairs (bp)]; (b) SCCHRIX (439,885 bp) and (c) YSCCHRVIN (270,148 bp). It is
also shown polynomial $P^{*}(l)\sim l^{\eta ^{\exp }}$ (solid line) and
exponential $P^{*}(l)\sim \exp (m^{\exp }l)$ (dashed line) fitting curves as
functions of exon size $l$. Results for $\eta ^{\exp }$ and $m^{\exp }$
obtained from these fittings in each case are in Table1.

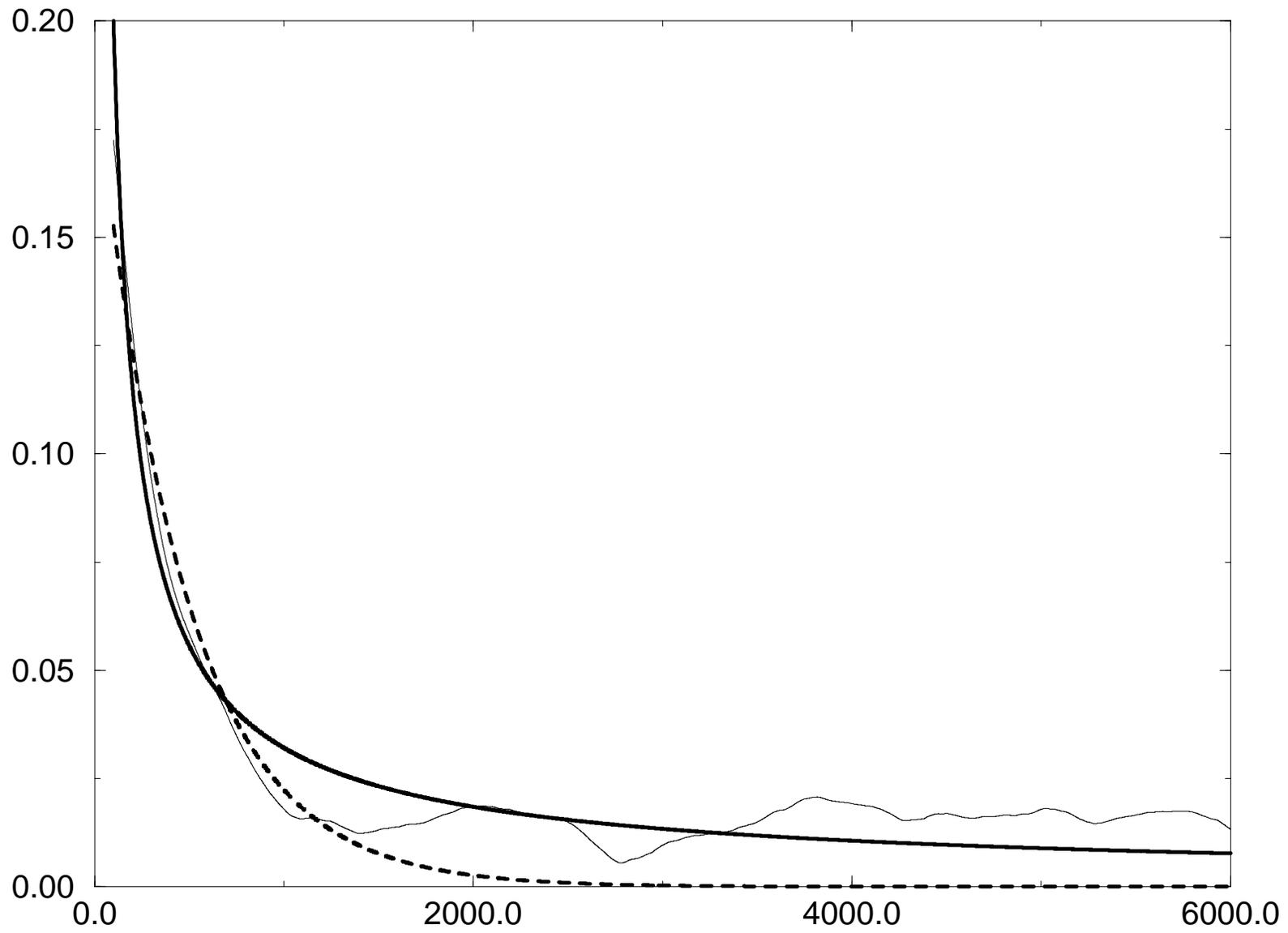
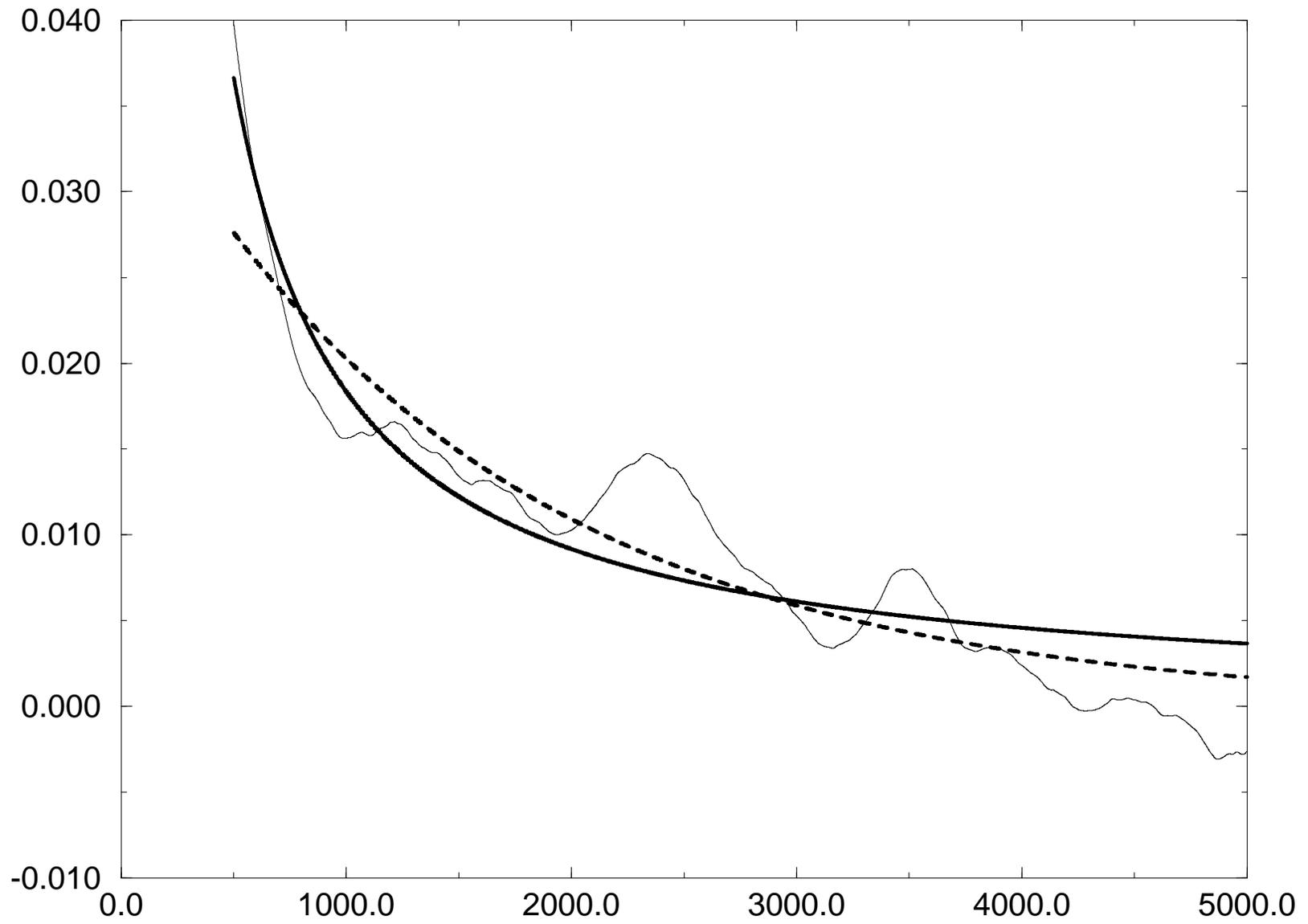
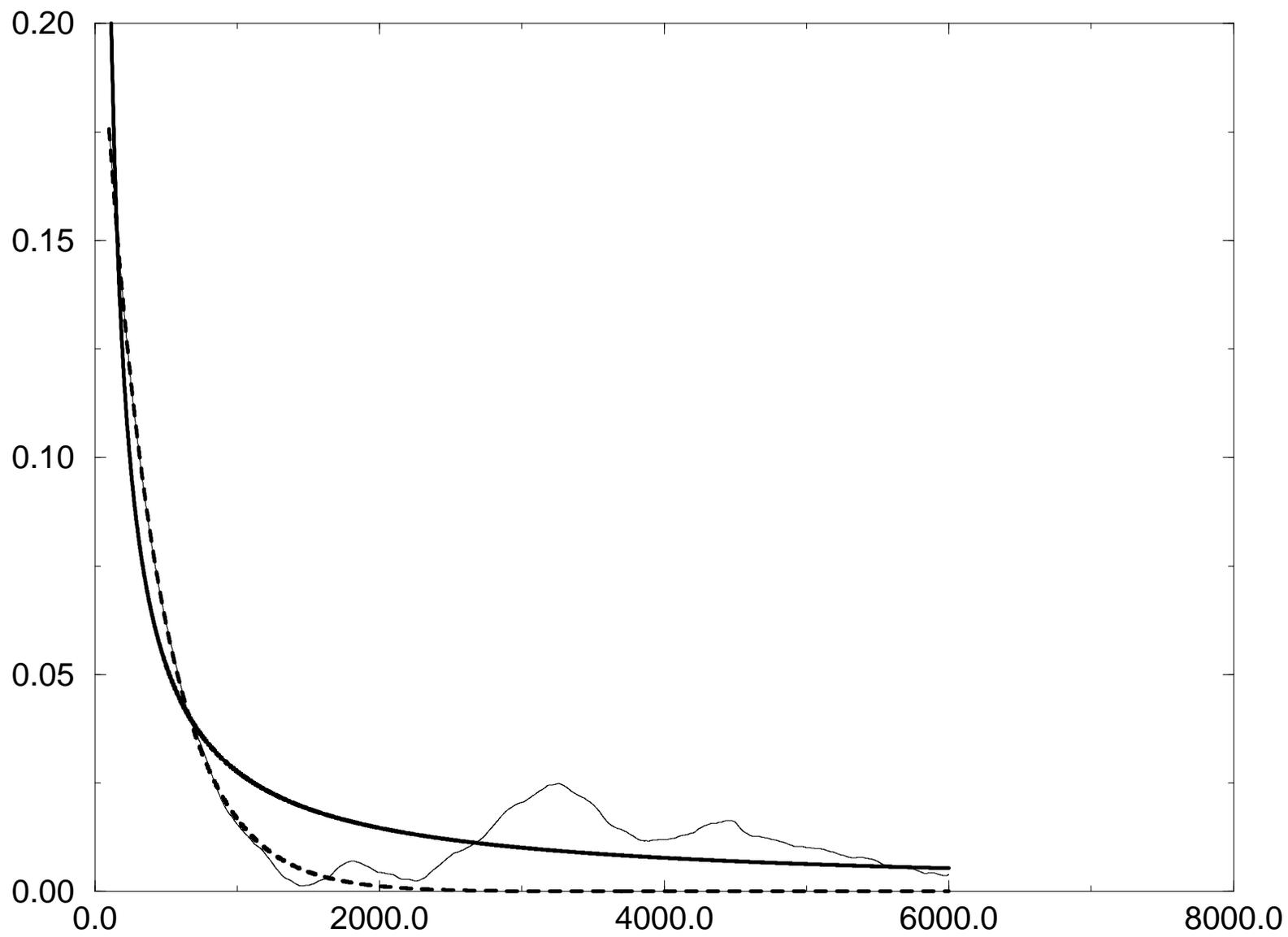
\item  {\bf Figure 3: }Corresponding{\bf \ }pair correlation functions $%
C^{*}(r)$ for the sequences of Figure 2. Polynomial $C^{*}(r)\sim
r^{\varepsilon ^{\exp }}$ and exponential $C^{*}(r)\sim \exp (n^{\exp }r)$
fitting curves are also depicted and results for $\varepsilon ^{\exp }$ and $%
n^{\exp }$ are in Table1.
\end{description}


\begin{thebibliography}{99}
\bibitem{exin}  - Berger et al, Proc. Natn. Acad. Sci. USA 74, 3171 (1977);

\bibitem{livrao}  - J.D. Watson, N.H. Hopkins, J.W. Roberts, J.A. Steitz,
A.M. Weiner, {\it Molecular Biology of the Gene}, fourth edition (The
Benjamin Cummings Publishing Company Inc., Ca 1987);

\bibitem{early}  - - W. Gilbert, Nature 271, 501 (1978);

\bibitem{late}  - W.F. Doolittle, Nature 272, 581 (1978);

\bibitem{proteinre}  - A. Stoltzfus, D.F. Spencer, M. Zuker, J.M. Logsdom
Jr., W.F. Doolittle , Science 265, 202 (1994);

\bibitem{origin}  - L.D. Hurst, Nature 371, 381 (1994);

\bibitem{stvosli}  - For earlier references on the subject, see W.Li,
Europhys. Lett. 10, 395 (1989); C.K. Peng, S.V. Buldyrev, A.L. Golberger, S.
Halvin, F. Scortino, M. Simons and H.E. Stanley, Nature 356, 168 (1992);
R.Voss, Phys. Rev. Lett 68, 3805 (1992); B. Borstnik, D. Pumpernik, D
Lukman, Europhy. Lett. 23 , 389 (1993);

\bibitem{nee}  - S. Nee, Nature 357, 450 (1992); S. Karlin and V. Brendel,
Science 259, 677 (1993); D. Larhammar and C.A. Chatzidimitrio- Dreissman,
Nucl. Acids Res. 21, 5167 (1993);

\bibitem{stanley93}  - S.V. Buldyrev, A.L. Goldberger, S. Havlin, C-K Peng,
M. Simons, H.E. Stanley, Phys. Rev E 47, 4514 (1993); Phys. Rev. E 49, 1685
(1994); Phys. Rev. Lett. 73, 3169 (1994); Phys. Rev E 51, 5084 (1995); W.Li,
T.G. Marr and K. Kaneko, Physica D 75, 392 (1994); A. Arneodo, E. Bacry,
P.V. Graves and J.F. Muzy, Phys.Rev.Lett 74, 3293 (1995); P.P.Galva\'an,
R.R. Rold\'an and J.L.Oliver, Phys. Rev E 53, 5181 (1996);

\bibitem{othermodel}  - These are based on Cellular Automata models. See for
example, H.Gutowitz, ed. Cellular Automata: Theory and Experiment, Physica
D45, 1990 (North Holland, Amsterdam, 1990); (MIT Press, Cambridge, MA, 1990);

\bibitem{eu}  - C. Goldman and A. Berezin, Phys. Rev. B 51, 12361 (1995);

\bibitem{fifel}  - B.U. Felderhof and M.E. Fisher, Ann of Phys. 58, 176
(1970); Ann of Phys. 58, 217 (1970); Ann. of Phys. 58, 268 (1970); Ann. of
Phys. 58, 281 (1970);

\bibitem{nota}  - The region of model parameters\thinspace considered in
Ref.[11] was restricted to those values for which $m_b<m_a.$ Extension of
the results for all regions can be easily obtained and the limit of $m_b\sim
3m_a$ can be taken.

\bibitem{codon}  - In biological literature a codon refers to a set of three
consecutive nucleotides on a DNA sequence that codifies for one specific
protein aminoacid.

\bibitem{isolated}  - According to Ref. \cite{eu}, an isolated {\it l}%
-cluster in the present context is defined as a set of {\it l} impurities
occupying consecutive sites of the lattice chain and separated from other
impurities at least by one host site at each end.

\bibitem{grad}  - I.S. Gradshteyn and I.M. Ryzhik, Table of Integrals,
Series and Products (Academic, San Diego, 1980).

\bibitem{fisher}  - M.E. Fisher, Physics 3, 255 (1967).

\bibitem{mosaic}  - In general, regions of nucleotides on an eukaryotic DNA
sequence comprise additional classification (e.g., flanking, operons, etc)
besides of being classified primarily as {\it exons} or {\it introns}. To
this fully structured sequence one reports to as the {\it mosaic structure }%
of DNA. The simplified view adopted here introduces no ambiguities and is
intended only to emphasize the multiscale characteristics of the genome with
respect to the coding and non-coding regions.

\bibitem{scale}  - The oscillatory behavior of correlation functions shall
be smoothed by accounting for an eventual degree of connectivity
(interaction) among the diverse coding regions along the considered
sequence; such interactions shall then preserve scale invariance of the
system. We should emphasize however that such a possibility has been totally
disregarded in our calculation.
\end{thebibliography}
\end{document}